\documentclass[fleqn,10pt]{wlpeerj} % for preprint submissions

\usepackage{lipsum}
\usepackage{longtable}

\usepackage{color}
\usepackage{fancyvrb}

\DefineVerbatimEnvironment{Highlighting}{Verbatim}{commandchars=\\\{\}}
\usepackage{framed}
\definecolor{shadecolor}{RGB}{248,248,248}
\newenvironment{Shaded}{\begin{snugshade}}{\end{snugshade}}

\newcommand{\CommentTok}[1]{\textcolor[rgb]{0.56,0.35,0.01}{\textit{#1}}}

\newcommand{\DataTypeTok}[1]{\textcolor[rgb]{0.13,0.29,0.53}{#1}}
\newcommand{\DecValTok}[1]{\textcolor[rgb]{0.00,0.00,0.81}{#1}}

\newcommand{\FloatTok}[1]{\textcolor[rgb]{0.00,0.00,0.81}{#1}}

\newcommand{\KeywordTok}[1]{\textcolor[rgb]{0.13,0.29,0.53}{\textbf{#1}}}
\newcommand{\NormalTok}[1]{#1}
\newcommand{\OperatorTok}[1]{\textcolor[rgb]{0.81,0.36,0.00}{\textbf{#1}}}
\newcommand{\OtherTok}[1]{\textcolor[rgb]{0.56,0.35,0.01}{#1}}

\newcommand{\StringTok}[1]{\textcolor[rgb]{0.31,0.60,0.02}{#1}}

\usepackage{booktabs}
\usepackage{array}
\usepackage{multirow}
\usepackage{wrapfig}
\usepackage{float}
\usepackage{colortbl}
\usepackage{pdflscape}
\usepackage{tabu}
\usepackage{threeparttable}
\usepackage{threeparttablex}
\usepackage[normalem]{ulem}
\usepackage{makecell}
\usepackage{xcolor}
\usepackage[utf8]{inputenc}
\usepackage{textcomp}
\usepackage{booktabs}
\usepackage{array}
\usepackage{multirow}
\usepackage{wrapfig}
\usepackage{float}
\usepackage{colortbl}
\usepackage{pdflscape}
\usepackage{tabu}
\usepackage{threeparttable}
\usepackage{threeparttablex}
\usepackage[normalem]{ulem}
\usepackage{makecell}
\usepackage{xcolor}

\title{Visualisation of Brain Statistics with R-packages \emph{ggseg} and \emph{ggseg3d}}

\author[1]{Athanasia M. Mowinckel}

\corrauthor[1]{Athanasia M. Mowinckel}{\href{mailto:a.m.mowinckel@psykologi.uio.no}{\nolinkurl{a.m.mowinckel@psykologi.uio.no}}}
\author[1]{Didac Vidal-Piñeiro}

\affil[1]{Center for Lifespan Changes in Brain and Cognition, University of Oslo, PO. box 1094 Blindern, 0317 Oslo, Norway}

\usepackage{natbib}
\begin{abstract}
There is an increased emphasis on visualizing neuroimaging results in more intuitive ways. Common statistical tools for dissemination, such as bar charts, lack the spatial dimension that is inherent in neuroimaging data. Here we present two packages for the statistical software R, \emph{ggseg} and \emph{ggseg3d}, that integrate this spatial component. The \emph{ggseg} and \emph{ggseg3d} packages visualize pre-defined brain segmentations as both 2D polygons and 3D meshes, respectively. Both packages are integrated with other well-established R-packages, allowing great flexibility. In this tutorial, we present the main data and functions in the \emph{ggseg} and \emph{ggseg3d} packages for brain atlas visualization. The main highlighted functions are able to display brain segmentation plots in R. Further, the accompanying \emph{ggsegExtra}-package includes a wider collection of atlases, and is intended for community-based efforts to develop more compatible atlases to \emph{ggseg} and \emph{ggseg3d}. Overall, the \emph{ggseg}-packages facilitate parcellation-based visualizations in R, improve and ease the dissemination of the results, and increase the efficiency of the workflows.
\end{abstract}

\begin{document}

\flushbottom
\maketitle
\thispagestyle{empty}

\hypertarget{introduction}{%
\section{Introduction}\label{introduction}}

Visualization is increasingly important for accurate guidance and interpretation of neuroimaging results, as current research is able to generate a high amount of data and outcomes.
For Magnetic Resonance Imaging (MRI), neuroimaging software provides whole-brain information by using many small units of space (\textgreater100.000).
Nonetheless, this data is often grouped and summarized into a limited number of regions using predefined brain parcellation atlases.
Brain parcellations segment the brain into a finite set of meaningful neurobiological components, which reflect one or more brain features either based on structural or connectivity properties (\citet{eickhoff_2018}).
The use of brain atlases is widespread as these facilitate interpretation and minimize the amount of data, hence reducing problems with multiple comparisons.
This enables replicability and data sharing in otherwise computationally expensive analyses, which are often performed in specialized software environments such as R (\citet{R}).

MRI data provides good spatial resolution and thus an optimal representation has to respect spatial relationships across regions.
Results from brain atlas analyses are most meaningfully visualized when projected onto a representation of the brain, thus it is desirable that any visual representation takes this relation into account.
The projection of data onto brain representations provides clear points of reference - especially when the reader is unfamiliar with the atlas - eases readability, guides interpretation, and conveys the spatial patterns of the data.
Adopting the grammar of graphics implemented in \emph{ggplot2} (\citet{ggplot}), one can plot neuroimaging data directly in R with several tools such as ggBrain (\citet{ggBrain}) and ggneuro (\citet{ggneuro}; see neuroconductor \citeyearpar{neuroconductor} for curated neuroimaging packages for R).
Yet these tools display whole-brain image files and are not well-suited for representing brain atlas data.

In this tutorial, we introduce two packages for visualizing brain atlas data in R.
The \emph{ggseg} and \emph{ggseg3d} -- pluss the complimentary \emph{ggsegExtra} -- packages include pre-compiled data sets for different brain atlases that allow for 2D and 3D visualization.
The two-dimensional functionality in \emph{ggseg} is based on polygons and \emph{ggplot2}-based grammar of graphics (\citet{ggplot}), while the 3D functionality in \emph{ggseg3d} is based on tri-surface mesh plots and \emph{plotly} (\citet{plotly}).

Both packages present compiled data sets, tailored functions that allow brain data integration and plotting, and other minor features such as custom colour palettes.
The data featured in the packages are derived from two well-known parcellations: the Desikan-Killany cortical atlas (DKT; \citet{dkt}), which covers the cortical surface of the brain, and the Automatic Segmentation of Subcortical Structures (aseg; \citet{aseg}), which covers the subcortical structures.
Both atlases are implemented in several neuroimaging softwares, such as FreeSurfer (\citet{fischl_99}, \citet{dale_99}, \citet{Fischl2000}), and are commonly used in relation to developmental changes, disease biomarkers, genomic data, and cognition (\citet{amlien_elaboration_2019}, \citet{WALHOVD20051261}, \citet{Pizzagalli}).
The \emph{ggsegExtra} package contains a collection of precompiled atlases (currently 15 additional atlases) and it is frequently updated.
A summary of all available atlases compatible with the \emph{ggseg}-packages as of December 2019 can be found in Table \ref{tab:atlasTab}.

\begin{table}

\caption{\label{tab:atlasTab}Table of currently available atlases in the ggseg, ggseg3d, and the ggsegExtra R-packages. Polygon and mesh refer to 2D and 3D brain atlas representations, respectively}
\centering
\resizebox{\linewidth}{!}{
\begin{tabular}[t]{lllll}
\toprule
Title & Item & Mesh & Polygon & citation\\
\midrule
Automated probabilistic reconstruction of white-matter pathways & tracula & None & ggsegExtra & \citet{tracula}\\
Desikan-Killiany Cortical Atlas & dkt & ggseg3d & ggseg & \citet{dkt}\\
Desterieux cortical parcellations & desterieux & ggsegExtra & None & \citet{desterieux}\\
Freesurfer automatic subcortical segmentation of a brain volume & aseg & ggseg3d & ggseg & \citet{aseg}\\
Genetic topography of brain area morphology & chenAr & None & ggsegExtra & \citet{chen}\\
\addlinespace
Genetic topography of brain thickness morphology & chenTh & None & ggsegExtra & \citet{chen}\\
Harvard-Oxford Cortical atlas & hoCort & None & ggsegExtra & \citet{ho}\\
HPC - Multi-modal parcellation of human cerebral cortex & glasser & None & ggsegExtra & \citet{glasser}\\
ICBM white matter parcellation & icbm & ggsegExtra & None & \citet{icbm}\\
JHU parcellation & jhu & None & ggsegExtra & \citet{jhu}\\
\addlinespace
Local-Global 17 Parcellation of the Human Cerebral Cortex & schaefer17 & ggsegExtra & None & \citet{schaefer}\\
Local-Global 7 Parcellation of the Human Cerebral Cortex & schaefer7 & ggsegExtra & None & \citet{schaefer}\\
Parcellation from JHU & jhu & ggsegExtra & None & \citet{jhu}\\
Parcellation from the Human Connectome Project & glasser & ggsegExtra & None & \citet{glasser}\\
White matter tract parcellations & tracula & ggsegExtra & None & \citet{tracula}\\
\addlinespace
Yeo 17 Resting-state Cortical Parcellations & yeo17 & ggsegExtra & ggsegExtra & \citet{yeo2011}\\
Yeo 7 Resting-state Cortical Parcellations & yeo7 & ggsegExtra & ggsegExtra & \citet{yeo2011}\\
\bottomrule
\end{tabular}}
\end{table}

\hypertarget{tutorial}{%
\section{Tutorial}\label{tutorial}}

This tutorial will introduce the \emph{ggseg}, \emph{ggseg3d}, and \emph{ggsegExtra} packages and familiarize the reader with the main functions and the general use of the packages.
The tutorial will focus on the two main functions: \texttt{ggseg()} for plotting 2D polygons and \texttt{ggseg3d()} for plotting 3D brains based on tri-surface mesh plots.

\hypertarget{plotting-polygon-data-ggplot2}{%
\subsection{Plotting polygon data (ggplot2)}\label{plotting-polygon-data-ggplot2}}

\begin{figure}
\centering
\includegraphics{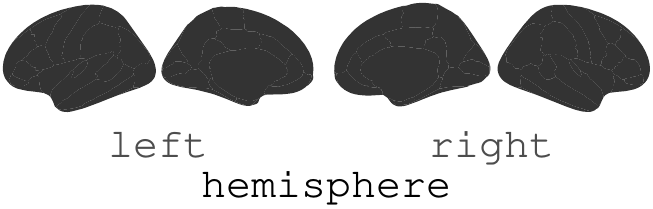}
\caption{\label{fig:init}By default \texttt{ggseg()} will plot the dkt atlas in grey shaded polygons.}
\end{figure}

\emph{ggseg} is the main function for plotting 2D data.
By default, the function automatically plots the DKT atlas (see Figure \ref{fig:init}).
The \texttt{ggseg()}-function is a wrapper for \texttt{geom\_polygon()} from \emph{ggplot2}, and it can be built upon and combined like any \emph{ggplot}-object.
The image plot consists of a simple brain representation containing no extra information.
Hence, \emph{ggseg} plots can be easily complemented with any of the available \emph{ggplot2} features and options.
We recommend users to get familiarized with \emph{ggplot2} (\citet{ggplot}).
The package is currently only available through github, but we expect to submit the \emph{ggseg}-package to The Comprehensive R Archive Network (\citet{cran}) in 2020.

\begin{Shaded}
\begin{Highlighting}[]
\CommentTok{\# remotes::install\_github("LCBC{-}UiO/ggseg")}
\KeywordTok{library}\NormalTok{(ggseg)}
\KeywordTok{library}\NormalTok{(dplyr)}
\KeywordTok{library}\NormalTok{(tidyr)}

\CommentTok{\# Figure 1}
\KeywordTok{ggseg}\NormalTok{()}
\end{Highlighting}
\end{Shaded}

\begin{figure}
\centering
\includegraphics{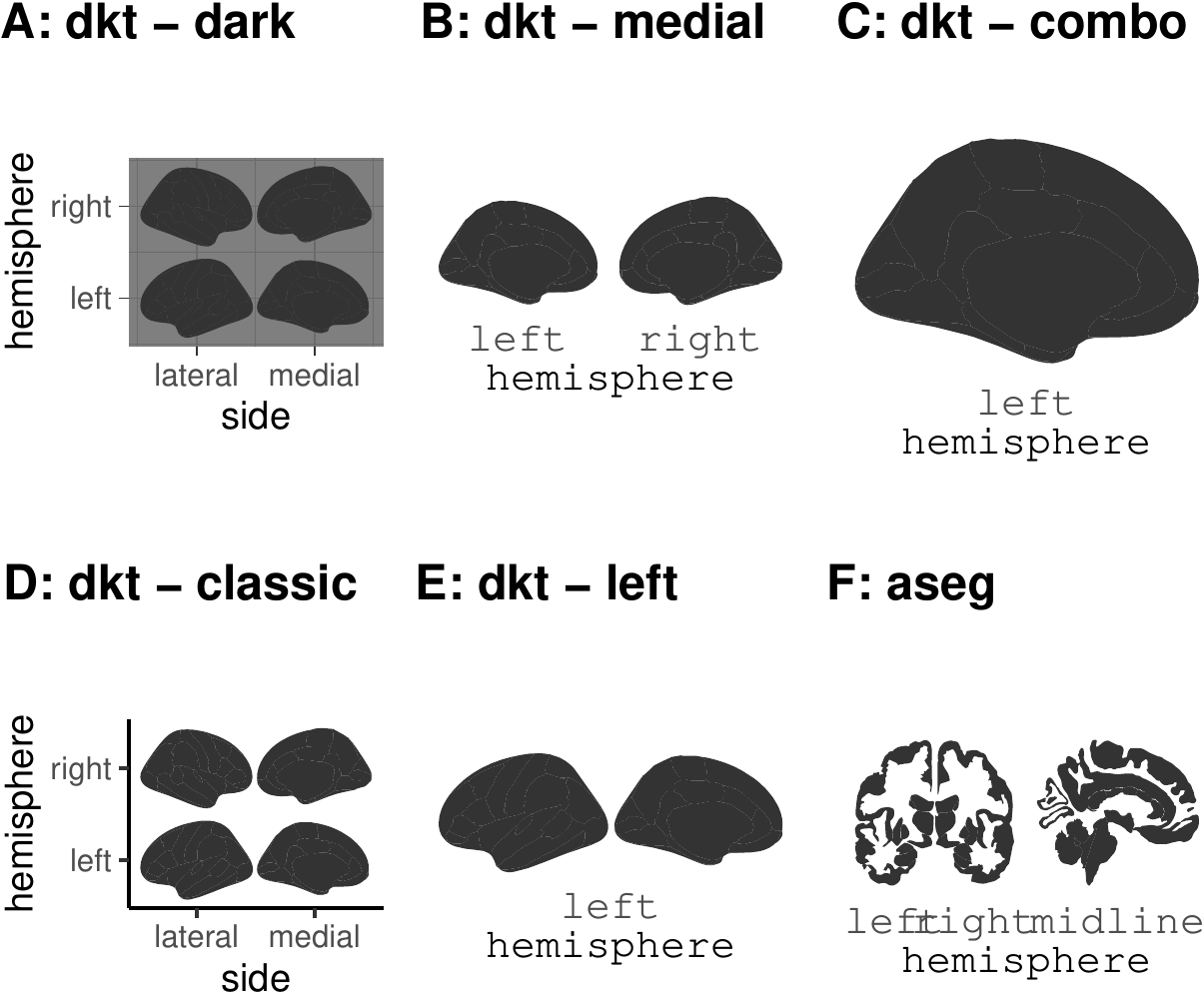}
\caption{\label{fig:collection}\emph{ggseg} plots can be used with any \emph{ggplot2} feature such as standard scales and themes. For cortical atlases, one can supply special \emph{ggseg} options to determine, for example, hemisphere, view, or position. \textbf{A:} \texttt{dkt} atlas, stacked with dark theme ; \textbf{B:} \texttt{dkt} with medial view only; \textbf{C:} \texttt{dkt} atlas with only left medial display; \textbf{D:} \texttt{dkt} atlas, stacked, with classic theme; \textbf{E:} \texttt{dkt} atlas with left hemisphere only; \textbf{F:} complete \texttt{aseg} atlas}
\end{figure}

In addition to the standard options for \emph{ggplot2} polygon geoms, the function also has several options for plotting the main brain representations.
These options are atlas-specific.
For cortical atlases, such as the \texttt{dkt}, one can stack the hemispheres, display only the medial or lateral side, choose either one or both hemispheres, or any combination of hemisphere and view (see Figure \ref{fig:collection} for examples).
For subcortical atlases, such as the \texttt{aseg} atlas, the options are more limited but one can often choose between axial, sagittal, and coronal views.

\begin{Shaded}
\begin{Highlighting}[]
\CommentTok{\# dkt dark theme}
\NormalTok{p1 <{-}}\StringTok{ }\KeywordTok{ggseg}\NormalTok{(}\DataTypeTok{position =} \StringTok{"stacked"}\NormalTok{) }\OperatorTok{+}
\StringTok{  }\KeywordTok{theme\_dark}\NormalTok{() }\OperatorTok{+}
\StringTok{  }\KeywordTok{labs}\NormalTok{(}\DataTypeTok{title=}\StringTok{" "}\NormalTok{)}

\CommentTok{\# dkt classic theme}
\NormalTok{p2 <{-}}\StringTok{ }\KeywordTok{ggseg}\NormalTok{(}\DataTypeTok{position =} \StringTok{"stacked"}\NormalTok{) }\OperatorTok{+}
\StringTok{  }\KeywordTok{theme\_classic}\NormalTok{() }\OperatorTok{+}
\StringTok{  }\KeywordTok{labs}\NormalTok{(}\DataTypeTok{title =} \StringTok{" "}\NormalTok{)}

\CommentTok{\# dkt medial view}
\NormalTok{med <{-}}\StringTok{ }\KeywordTok{ggseg}\NormalTok{(}\DataTypeTok{view =} \StringTok{"medial"}\NormalTok{) }\OperatorTok{+}
\StringTok{  }\KeywordTok{labs}\NormalTok{(}\DataTypeTok{title =} \StringTok{" "}\NormalTok{)}

\CommentTok{\# dkt left hemisphere}
\NormalTok{left <{-}}\StringTok{ }\KeywordTok{ggseg}\NormalTok{(}\DataTypeTok{hemisphere =} \StringTok{"left"}\NormalTok{) }\OperatorTok{+}
\StringTok{  }\KeywordTok{labs}\NormalTok{(}\DataTypeTok{title =} \StringTok{" "}\NormalTok{)}

\CommentTok{\# aseg default theme}
\NormalTok{p3 <{-}}\StringTok{ }\KeywordTok{ggseg}\NormalTok{(}\DataTypeTok{atlas=}\NormalTok{aseg) }\OperatorTok{+}
\StringTok{  }\KeywordTok{labs}\NormalTok{(}\DataTypeTok{title =} \StringTok{" "}\NormalTok{)}

\CommentTok{\# dkt left medial alone}
\NormalTok{combo <{-}}\StringTok{ }\KeywordTok{ggseg}\NormalTok{(}\DataTypeTok{view =} \StringTok{"medial"}\NormalTok{,}
               \DataTypeTok{hemisphere =} \StringTok{"left"}\NormalTok{) }\OperatorTok{+}
\StringTok{  }\KeywordTok{labs}\NormalTok{(}\DataTypeTok{title=}\StringTok{" "}\NormalTok{)}

\CommentTok{\# Combine plots to Figure 2}
\NormalTok{cowplot}\OperatorTok{::}\KeywordTok{plot\_grid}\NormalTok{(p1, med, combo,  p2, left, p3,}
                   \DataTypeTok{labels =} \KeywordTok{c}\NormalTok{(}\StringTok{"A: dkt {-} dark"}\NormalTok{, }\StringTok{"B: dkt {-} medial"}\NormalTok{,}
                              \StringTok{"C: dkt {-} combo"}\NormalTok{, }\StringTok{"D: dkt {-} classic"}\NormalTok{,}
                              \StringTok{"E: dkt {-} left"}\NormalTok{, }\StringTok{"F: aseg"}\NormalTok{),}
                   \DataTypeTok{hjust =} \FloatTok{{-}.05}\NormalTok{)}
\end{Highlighting}
\end{Shaded}

\hypertarget{using-own-data-with-fill-and-colour}{%
\subsubsection{Using own data with fill and colour}\label{using-own-data-with-fill-and-colour}}

\begin{figure}
\centering
\includegraphics{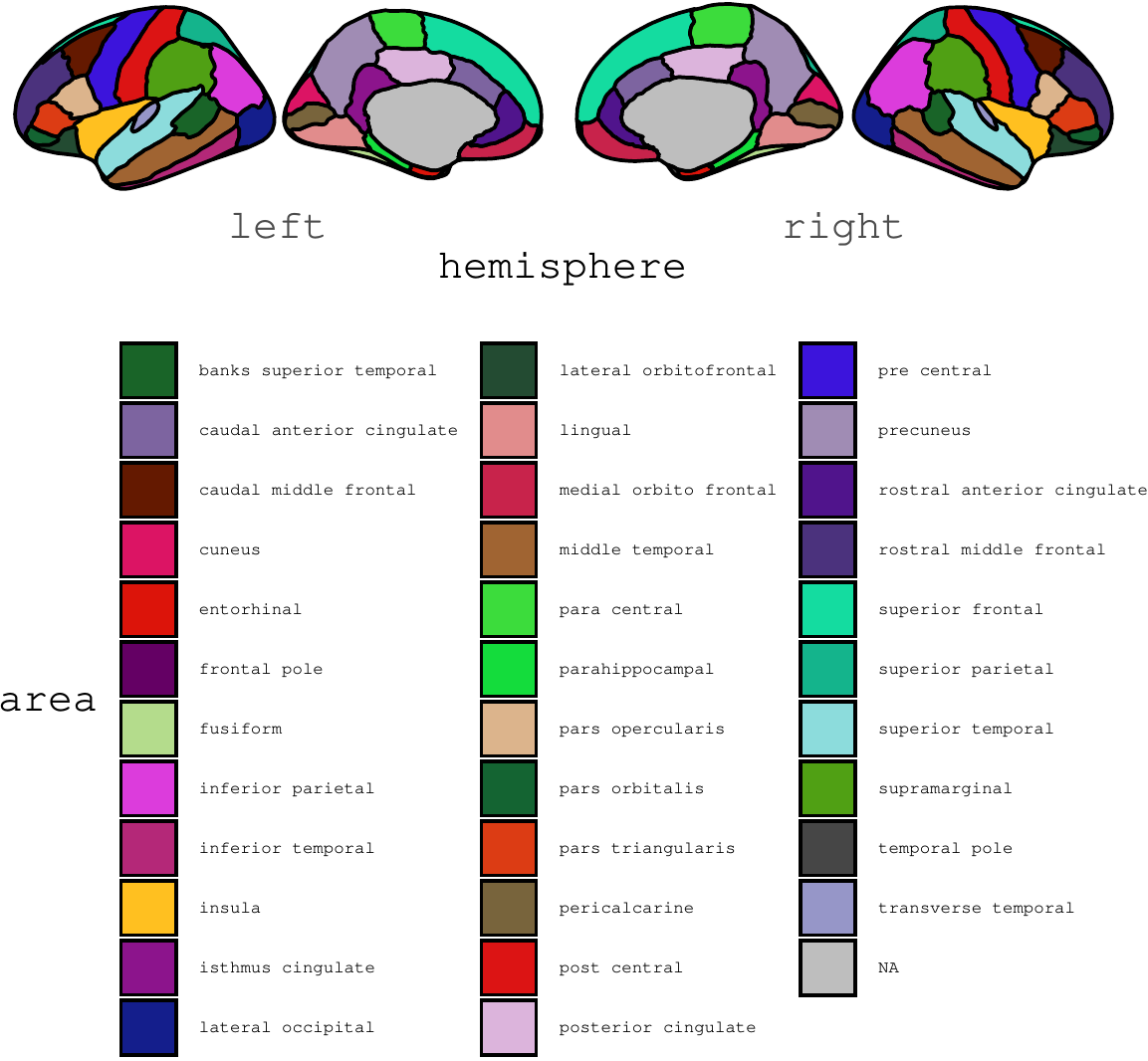}
\caption{\label{fig:fill}Supplying `area' to the fill option in \texttt{ggseg()}, will use the column `area' from the accompanying dataset to create a discrete colour palette over the segments in the atlas. The \texttt{dkt} atlas default palette corresponds to the FreeSurferColorLut scheme.}
\end{figure}

\texttt{ggseg()} accepts any argument you can supply to \texttt{geom\_polygon()}, and therefore is easy to work with for those familiar with \emph{ggplot2} functionality.
Standard arguments like \texttt{fill} that floods the segments with a colour, or \texttt{colour} that colours the edges around the segments are typical arguments to provide to the function either as a single setting value or within the \emph{ggplot2} mapping function \texttt{aes}.
To use color palettes corresponding to those used in the original neuroimaging softwares one can use atlas-specific `brain' palette scales (Figure \ref{fig:fill}).

\begin{Shaded}
\begin{Highlighting}[]
\CommentTok{\# Figure 3}
\KeywordTok{ggseg}\NormalTok{(}\DataTypeTok{mapping=}\KeywordTok{aes}\NormalTok{(}\DataTypeTok{fill =}\NormalTok{ area), }\DataTypeTok{colour=}\StringTok{"black"}\NormalTok{) }\OperatorTok{+}
\StringTok{  }\KeywordTok{scale\_fill\_brain}\NormalTok{(}\StringTok{"dkt"}\NormalTok{) }\OperatorTok{+}
\StringTok{  }\KeywordTok{theme}\NormalTok{(}\DataTypeTok{legend.justification=}\KeywordTok{c}\NormalTok{(}\DecValTok{1}\NormalTok{,}\DecValTok{0}\NormalTok{),}
        \DataTypeTok{legend.position=}\StringTok{"bottom"}\NormalTok{,}
        \DataTypeTok{legend.text =} \KeywordTok{element\_text}\NormalTok{(}\DataTypeTok{size =} \DecValTok{5}\NormalTok{)) }\OperatorTok{+}
\StringTok{  }\KeywordTok{guides}\NormalTok{(}\DataTypeTok{fill =} \KeywordTok{guide\_legend}\NormalTok{(}\DataTypeTok{ncol =} \DecValTok{3}\NormalTok{))}
\end{Highlighting}
\end{Shaded}

Most users will use \texttt{ggseg()} to display - using a color scale - some descriptive or inferential statistics, such as mean thickness or brain-cognition relationships across the different brain regions.
Yet, before projecting the statistics onto the segments, one should explore the structure of the atlas data sets.
The atlas data set structure will help users understand what incoming statistical data needs to look like.
Note that each atlas corresponds to a unique data set.
All data sets have a similar structure and contain key information regarding the atlas, the region names, and the coordinates for the segment polygons.

\small

\begin{Shaded}
\begin{Highlighting}[]
\CommentTok{\# Look at the top 5 rows of the dkt dataset}
\KeywordTok{head}\NormalTok{(dkt, }\DecValTok{5}\NormalTok{)}
\end{Highlighting}
\end{Shaded}

\normalsize

In any atlas, the column `label' is particularly useful for combining the data of interest with the \emph{ggseg}-polygons.
The column `label' contains the label (region) names as in the original neuroimaging software.
For example, the DKT atlas label column matches the region names from Freesurfer statistics table outputs.
Yet, the data in \emph{ggseg} is in a long format - that is each region has a single row - and any data of interest needs to be in this same format.
Often data sets are organized in wide format, in which subjects are represented by rows and each different data variable is represented in a separate column, and thus need to be rearranged to work with ggseg.
See below an example of wide-to-long conversion.

\small

\begin{Shaded}
\begin{Highlighting}[]
\CommentTok{\# Create a mock freesurfer output}
\NormalTok{freesurfer\_stats <{-}}\StringTok{ }\KeywordTok{data.frame}\NormalTok{(}
  \DataTypeTok{id =} \KeywordTok{c}\NormalTok{(}\DecValTok{10}\OperatorTok{:}\DecValTok{12}\NormalTok{),}
  \DataTypeTok{lh\_superiortemporal =} \KeywordTok{c}\NormalTok{(}\FloatTok{3.32}\NormalTok{, }\FloatTok{4.1}\NormalTok{, }\FloatTok{3.5}\NormalTok{),}
  \DataTypeTok{lh\_precentral =} \KeywordTok{c}\NormalTok{(}\FloatTok{2.3}\NormalTok{, }\FloatTok{2.5}\NormalTok{, }\FloatTok{2.1}\NormalTok{),}
  \DataTypeTok{lh\_rostralmiddlefrontal =} \KeywordTok{c}\NormalTok{(}\FloatTok{3.3}\NormalTok{, }\FloatTok{3.2}\NormalTok{, }\FloatTok{3.1}\NormalTok{)}
\NormalTok{)}
\NormalTok{freesurfer\_stats}

\CommentTok{\# Wrangle wide format data into long format}
\NormalTok{freesurfer\_long <{-}}\StringTok{ }\NormalTok{freesurfer\_stats }\OperatorTok{\%>\%}
\StringTok{  }\KeywordTok{gather}\NormalTok{(label, thickness, }\OperatorTok{{-}}\NormalTok{id)}
\NormalTok{freesurfer\_long}
\CommentTok{\#\#   id lh\_superiortemporal lh\_precentral lh\_rostralmiddlefrontal}
\CommentTok{\#\# 1 10                3.32           2.3                     3.3}
\CommentTok{\#\# 2 11                4.10           2.5                     3.2}
\CommentTok{\#\# 3 12                3.50           2.1                     3.1}
\CommentTok{\#\#   id                   label thickness}
\CommentTok{\#\# 1 10     lh\_superiortemporal      3.32}
\CommentTok{\#\# 2 11     lh\_superiortemporal      4.10}
\CommentTok{\#\# 3 12     lh\_superiortemporal      3.50}
\CommentTok{\#\# 4 10           lh\_precentral      2.30}
\CommentTok{\#\# 5 11           lh\_precentral      2.50}
\CommentTok{\#\# 6 12           lh\_precentral      2.10}
\CommentTok{\#\# 7 10 lh\_rostralmiddlefrontal      3.30}
\CommentTok{\#\# 8 11 lh\_rostralmiddlefrontal      3.20}
\CommentTok{\#\# 9 12 lh\_rostralmiddlefrontal      3.10}
\end{Highlighting}
\end{Shaded}

\normalsize

\begin{figure}
\centering
\includegraphics{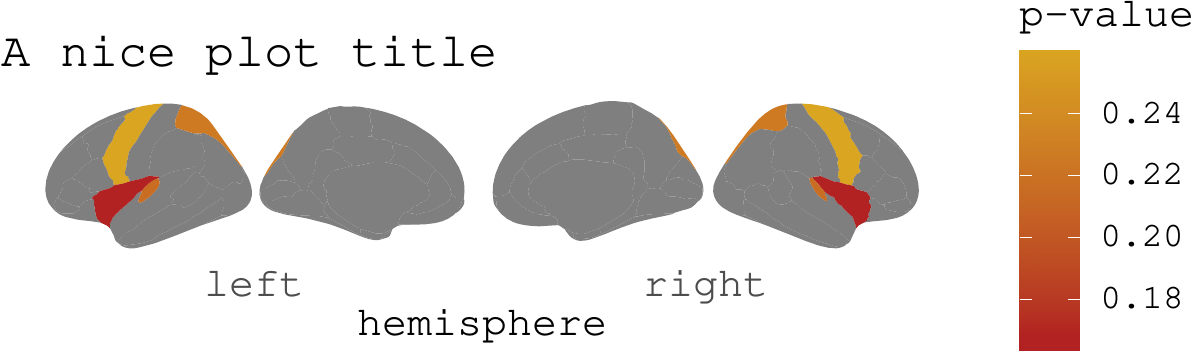}
\caption{\label{fig:datasupp}Supplying data through the `.data' option in \texttt{ggseg()} enables use of columns in the supplied data to aesthetical arguments (such as `fill'). The \emph{ggseg} plot can be used with any other polygon compatible function from \emph{ggplot2} or \emph{ggplot2} extentions, for instance adding title, changing the legend name and the colour scheme with standard \emph{ggplot2} functions.}
\end{figure}

Data in long format can then be used directly with the \texttt{ggseg()}-function, as the `label' column corresponds in name and content with the `label' column in the atlas data of \texttt{dkt}.
The data \textbf{must} include a column that has the same column name and at least \emph{some} data matching the values in the corresponding column in the atlas data.
In the next example we create some data with 4 rows, with an `area' and `p' column, representing the results of a hypothetical analysis.
The \texttt{ggseg()}-function will recognise the matching column `area', and merge the supplied data into the atlas using \emph{dplyr}-joins.
We use the `p' column as the column flooding the segment with colour.
The appearance of the plot can then be modified similarly to any other \emph{ggplot2} graph using functions such as scales, labs, themes, etc., as seen in Figure \ref{fig:datasupp}

\begin{Shaded}
\begin{Highlighting}[]
\CommentTok{\# Make some mock data}
\NormalTok{someData =}\StringTok{ }\KeywordTok{data.frame}\NormalTok{(}
  \DataTypeTok{area =} \KeywordTok{c}\NormalTok{(}\StringTok{"transverse temporal"}\NormalTok{, }\StringTok{"insula"}\NormalTok{,}
           \StringTok{"pre central"}\NormalTok{,}\StringTok{"superior parietal"}\NormalTok{),}
  \DataTypeTok{p =} \KeywordTok{sample}\NormalTok{(}\KeywordTok{seq}\NormalTok{(}\DecValTok{0}\NormalTok{,.}\DecValTok{5}\NormalTok{,.}\DecValTok{001}\NormalTok{), }\DecValTok{4}\NormalTok{),}
  \DataTypeTok{stringsAsFactors =} \OtherTok{FALSE}\NormalTok{)}

\CommentTok{\# Figure 4}
\KeywordTok{ggseg}\NormalTok{(}\DataTypeTok{.data=}\NormalTok{someData, }\DataTypeTok{mapping=}\KeywordTok{aes}\NormalTok{(}\DataTypeTok{fill=}\NormalTok{p))  }\OperatorTok{+}
\StringTok{  }\KeywordTok{labs}\NormalTok{(}\DataTypeTok{title=}\StringTok{"A nice plot title"}\NormalTok{, }\DataTypeTok{fill=}\StringTok{"p{-}value"}\NormalTok{) }\OperatorTok{+}
\StringTok{  }\KeywordTok{scale\_fill\_gradient}\NormalTok{(}\DataTypeTok{low=}\StringTok{"firebrick"}\NormalTok{,}\DataTypeTok{high=}\StringTok{"goldenrod"}\NormalTok{)}
\end{Highlighting}
\end{Shaded}

If the results are only in one hemisphere, but you still want to plot both of them, make sure your data.frame includes the column `hemi' with either `right' or `left'.
In this case, data will be merged into the atlas both by `area' and by `hemi'.
For more information about adapting data and viewing only one hemisphere or side, the \href{https://lcbc-uio.github.io/ggseg/articles/ggseg.html\#single-hemisphere-results}{package vignettes} contains more elaborate information.

\hypertarget{creating-subplots}{%
\subsubsection{Creating subplots}\label{creating-subplots}}

\begin{figure}
\centering
\includegraphics{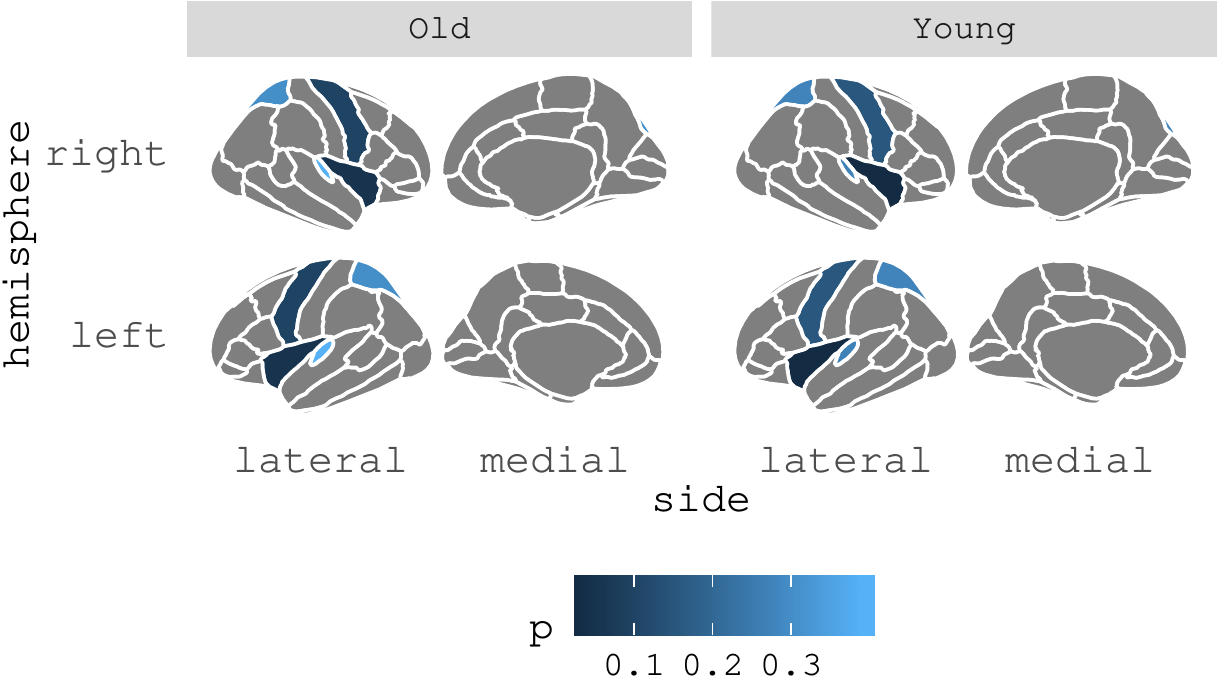}
\caption{\label{fig:datasupp3}To visualize data corresponding to different groups, one needs to use \texttt{group\_by} or similar \textbf{before} plotting with \texttt{ggseg()} for facetting to work.}
\end{figure}

There is often the need to plot a statistic of interest in different groups (e.g.~thickness or brain - cognition relationships in young or older adults).
This may be obtained also with \texttt{ggseg()}, using \emph{ggplot2}'s \texttt{facet\_wrap} or \texttt{facet\_grid}, with three guiding rules:
\textbf{1)} as before, data needs to be in long format with a column indexing which group the row corresponds to (group data should appear in seperate rows, not in separate columns).
\textbf{2)} The data needs to be grouped using \emph{dplyr}'s \texttt{group\_by()} function \emph{before} providing the data to the \texttt{ggseg()}-function.
The \texttt{ggseg()}-function will detect grouped data, and adapt it to \texttt{facet}`s requirements.
\textbf{3)} Apply \texttt{facet\_wrap} or \texttt{facet\_grid} to the plot having used the above two rules.
An example of this can be seen in Figure \ref{fig:datasupp3}, where a mock data set including summary statistics for two groups ('Young' and `Old') is used when faceting a \emph{ggseg}-plot.

\begin{figure}
\centering
\includegraphics{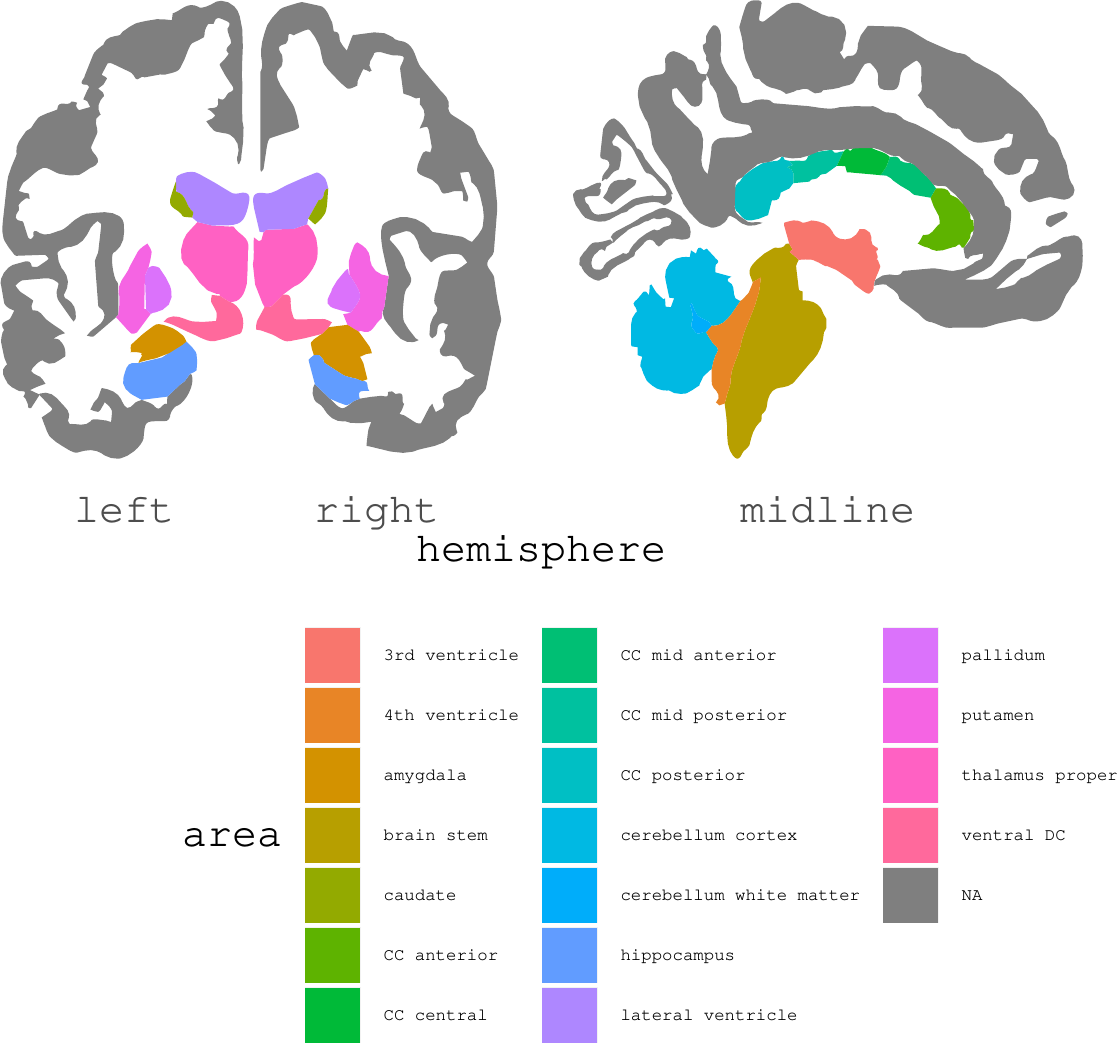}
\caption{\label{fig:atlases}The \texttt{aseg} atlas showing subcortical structures has some distinct differences from the \texttt{dkt}. For instance, there is no option to show only a single hemisphere. Furthermore, rather than showing lateral and medial surfaces, it shows an axial and sagittal slice.}
\end{figure}

All the concepts described above also work with the \texttt{aseg} atlas for subcortical structures, except for `hemisphere' and `view' arguments that are superfluous in subcortical atlases (Figure \ref{fig:atlases}).

\begin{Shaded}
\begin{Highlighting}[]
\CommentTok{\# Make some mock data}
\NormalTok{someData =}\StringTok{ }\KeywordTok{data.frame}\NormalTok{(}
  \DataTypeTok{area =} \KeywordTok{rep}\NormalTok{(}\KeywordTok{c}\NormalTok{(}\StringTok{"transverse temporal"}\NormalTok{, }\StringTok{"insula"}\NormalTok{,}
               \StringTok{"pre central"}\NormalTok{,}\StringTok{"superior parietal"}\NormalTok{),}\DecValTok{2}\NormalTok{),}
  \DataTypeTok{p =} \KeywordTok{sample}\NormalTok{(}\KeywordTok{seq}\NormalTok{(}\DecValTok{0}\NormalTok{,.}\DecValTok{5}\NormalTok{,.}\DecValTok{001}\NormalTok{), }\DecValTok{8}\NormalTok{),}
  \DataTypeTok{AgeG =} \KeywordTok{c}\NormalTok{(}\KeywordTok{rep}\NormalTok{(}\StringTok{"Young"}\NormalTok{,}\DecValTok{4}\NormalTok{), }\KeywordTok{rep}\NormalTok{(}\StringTok{"Old"}\NormalTok{,}\DecValTok{4}\NormalTok{)),}
  \DataTypeTok{stringsAsFactors =} \OtherTok{FALSE}\NormalTok{) }\OperatorTok{\%>\%}
\StringTok{  }\KeywordTok{group\_by}\NormalTok{(AgeG)}

\CommentTok{\# Figure 5}
\KeywordTok{ggseg}\NormalTok{(}\DataTypeTok{.data=}\NormalTok{someData, }\DataTypeTok{colour=}\StringTok{"white"}\NormalTok{, }\DataTypeTok{position =} \StringTok{"stacked"}\NormalTok{,}
      \DataTypeTok{mapping=}\KeywordTok{aes}\NormalTok{(}\DataTypeTok{fill=}\NormalTok{p)) }\OperatorTok{+}
\StringTok{  }\KeywordTok{facet\_wrap}\NormalTok{(}\OperatorTok{\textasciitilde{}}\NormalTok{AgeG, }\DataTypeTok{ncol=}\DecValTok{2}\NormalTok{) }\OperatorTok{+}
\StringTok{  }\KeywordTok{theme}\NormalTok{(}\DataTypeTok{legend.position =} \StringTok{"bottom"}\NormalTok{)}
\end{Highlighting}
\end{Shaded}

\begin{Shaded}
\begin{Highlighting}[]
\CommentTok{\# Figure 6}
\KeywordTok{ggseg}\NormalTok{(}\DataTypeTok{atlas=}\StringTok{"aseg"}\NormalTok{, }\DataTypeTok{mapping=}\KeywordTok{aes}\NormalTok{(}\DataTypeTok{fill=}\NormalTok{area)) }\OperatorTok{+}\StringTok{ }
\StringTok{  }\KeywordTok{theme}\NormalTok{(}\DataTypeTok{legend.justification=}\KeywordTok{c}\NormalTok{(}\DecValTok{1}\NormalTok{,}\DecValTok{0}\NormalTok{),}
        \DataTypeTok{legend.position=}\StringTok{"bottom"}\NormalTok{,}
        \DataTypeTok{legend.text =} \KeywordTok{element\_text}\NormalTok{(}\DataTypeTok{size =} \DecValTok{5}\NormalTok{)) }\OperatorTok{+}
\StringTok{  }\KeywordTok{guides}\NormalTok{(}\DataTypeTok{fill =} \KeywordTok{guide\_legend}\NormalTok{(}\DataTypeTok{ncol =} \DecValTok{3}\NormalTok{))}
\end{Highlighting}
\end{Shaded}

\hypertarget{plotting-3d-mesh-data}{%
\subsection{Plotting 3D mesh data}\label{plotting-3d-mesh-data}}

Representing brains as 2D polygons is a good solution for fast, efficient, and flexible plotting, and can be easily combined with interactive apps such as Shiny (\citet{shiny}).
Yet, brains are intrinsically 3-dimensional and it can be challenging to recognize the location of a region as a flattened image.
This problem is exacerbated in atlases that represent subcortical features as they are 3-dimensional, while cortical structures, such as grey matter structures, can be flattened to 2-dimensions.
Hence, here we also provide the \emph{ggseg3d} package to plot, view, and print 3D-atlases in R.
\emph{ggseg3d} is based on tri-surface mesh plots using \emph{plotly} (\citet{plotly}).
The data structure is more complex than the \emph{ggplot2} polygons, and includes additional options for brain inflation, glass brains, camera locations, etc.
As \emph{ggseg3d} is based on plotly, the resulting brain atlases are interactive, which guides interpretation, and is useful for public dissemination.
We recommend users to familiarize themselves with \emph{plotly} (\citet{plotly}) when using this function.

\begin{figure}[H]
\includegraphics[width=0.3\linewidth]{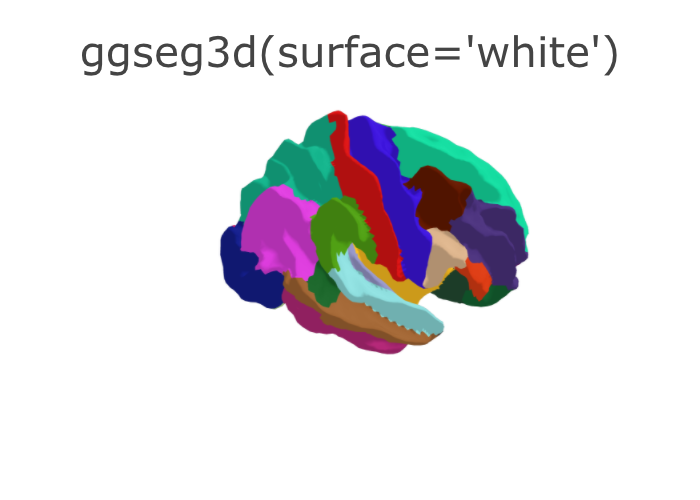} \includegraphics[width=0.3\linewidth]{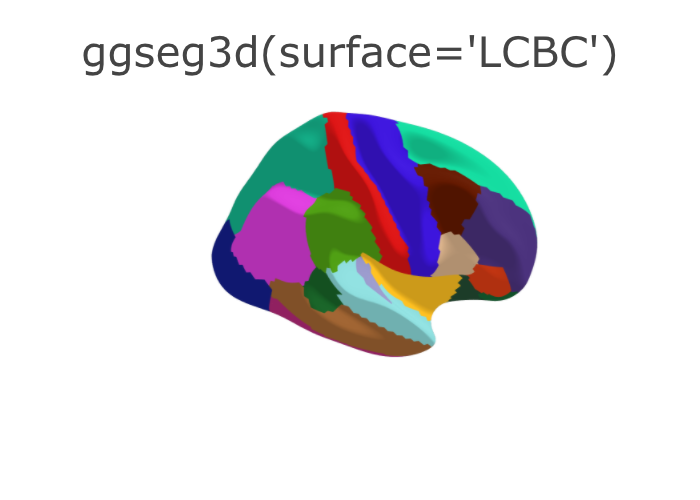} \includegraphics[width=0.3\linewidth]{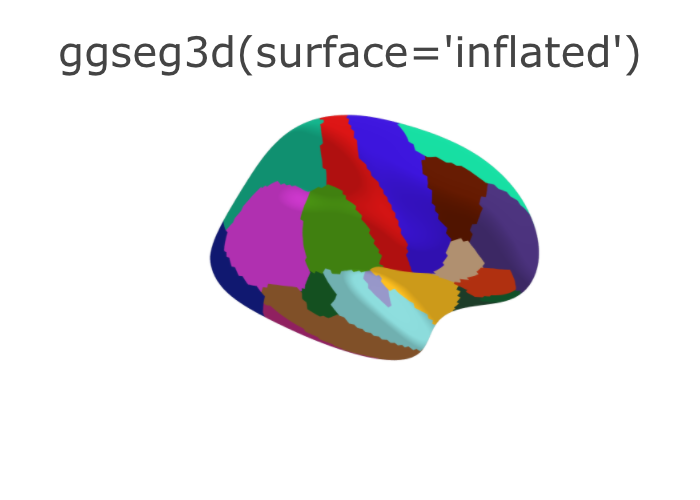} \caption{The three surface options provided in \emph{ggseg3d} atlases. \textbf{From left to right:} the `white' surface is the white matter surface, `LCBC' surface is the semi-inflated white matter surface (inflated over 10 iterations), and the `inflated' surface is a inflated grey matter cortical surface as provided by the FreeSurfer software.}\label{fig:ggseg3d-1-out}
\end{figure}

Out-of-the-box, \texttt{ggseg3d()} plots the \texttt{dkt\_3d} atlas in `LCBC' surface, but there are two more surfaces available for cortical atlases (Figure \ref{fig:ggseg3d-1-out})
The `LCBC' surface consists on a semi-inflated white matter surface based on the \emph{fsaverage5} template subject.
All \texttt{{[}...{]}\_3d} atlases include a \texttt{colour} column that based on the color scheme used in the source neuroimaging software.

The 3D-atlas data is stored in nested tibbles.
Each cortical atlas has data sets for three different surfaces (see Figure \ref{fig:ggseg3d-1-out}) and the two hemispheres.
Only one surface is available for subcortical atlases as inflation procedures are irrelevant.
The `ggseg\_3d' column includes all necessary information for \texttt{ggseg3d()} to create a 3D mesh-plot, and should not be modified by the user.
The additional 3D-atlases in \emph{ggsegExtra} have the same data structure.
It is important to note that the coordinates in the plot (X, Y, Z) are \textbf{not} any type of radiological coordinate system, but arbitrary Cartesian plot coordinates.

\small

\begin{Shaded}
\begin{Highlighting}[]
\CommentTok{\# remotes::install\_github("LCBC{-}UiO/ggseg3d")}
\KeywordTok{library}\NormalTok{(ggseg3d)}
\end{Highlighting}
\end{Shaded}

\normalsize

\hypertarget{external-data-supply}{%
\subsubsection{External data supply}\label{external-data-supply}}

Similarly as in the 2D-atlas, the user will use \texttt{ggseg3d()} to display through a colour scale some descriptive or inferential statistic.
If the data is not already in the correct long format or uses similar naming as the atlas, the users should inspect the atlas data for a specific surface (and hemisphere, if desired), and then \texttt{unnest(ggseg\_3d)} it to see how the atlas data is organised.

\small

\begin{Shaded}
\begin{Highlighting}[]
\CommentTok{\# Select surface and hemisphere, and then unnest to inspect the atlas data}
\NormalTok{dkt\_3d }\OperatorTok{\%>\%}
\StringTok{  }\KeywordTok{filter}\NormalTok{(surf }\OperatorTok{==}\StringTok{ "inflated"} \OperatorTok{\&}\StringTok{ }\NormalTok{hemi }\OperatorTok{==}\StringTok{ "right"}\NormalTok{) }\OperatorTok{\%>\%}
\StringTok{  }\KeywordTok{unnest}\NormalTok{(ggseg\_3d) }\OperatorTok{\%>\%}
\StringTok{  }\KeywordTok{head}\NormalTok{(}\DecValTok{5}\NormalTok{)}
\end{Highlighting}
\end{Shaded}

\normalsize

Note the \texttt{mesh} column, which contains lists.
Each list corresponds to a region and contains 6 vectors required to create the mesh of the tri-surface plot.
It should also be noted that the `label', `annot' and `area' columns could provide matching values for your own data.
Similarly to the \texttt{ggseg()}-function, the `label' column should match the region names used in the original neuroimaging software while `area' and `annot' provide alternative/secondary names.
It is thus important to match your regional identifiers with those used in the atlas.
To colour the segments using a column from the data, a column name from the data needs to be supplied to the \texttt{colour} option, and providing it to the \texttt{text} option will add another line to the \emph{plotly} hover information.

\begin{Shaded}
\begin{Highlighting}[]
\CommentTok{\# Figure 8 left}
\KeywordTok{ggseg3d}\NormalTok{(}\DataTypeTok{.data =}\NormalTok{ someData, }\DataTypeTok{atlas =}\NormalTok{ dkt\_3d, }\DataTypeTok{colour =} \StringTok{"p"}\NormalTok{, }\DataTypeTok{text =} \StringTok{"p"}\NormalTok{)}

\CommentTok{\# Figure 8 middle}
\KeywordTok{ggseg3d}\NormalTok{(}\DataTypeTok{.data =}\NormalTok{ someData, }
        \DataTypeTok{atlas =}\NormalTok{ dkt\_3d,}
        \DataTypeTok{colour =} \StringTok{"p"}\NormalTok{, }\DataTypeTok{text =} \StringTok{"p"}\NormalTok{,}
        \DataTypeTok{palette =} \KeywordTok{c}\NormalTok{(}\StringTok{"\#ff0000"}\NormalTok{, }\StringTok{"\#00ff00"}\NormalTok{, }\StringTok{"\#0000ff"}\NormalTok{)) }\OperatorTok{\%>\%}\StringTok{   }
\StringTok{  }\KeywordTok{pan\_camera}\NormalTok{(}\StringTok{"right lateral"}\NormalTok{) }\OperatorTok{\%>\%}\StringTok{ }
\StringTok{  }\KeywordTok{remove\_axes}\NormalTok{()}

\CommentTok{\# Figure 8 right}
\KeywordTok{ggseg3d}\NormalTok{(}\DataTypeTok{.data =}\NormalTok{ someData, }
        \DataTypeTok{atlas =}\NormalTok{ dkt\_3d,}
        \DataTypeTok{colour =} \StringTok{"p"}\NormalTok{, }\DataTypeTok{text =} \StringTok{"p"}\NormalTok{,}
        \DataTypeTok{na.colour =} \StringTok{"black"}\NormalTok{) }\OperatorTok{\%>\%}\StringTok{   }
\StringTok{  }\KeywordTok{pan\_camera}\NormalTok{(}\StringTok{"right lateral"}\NormalTok{) }\OperatorTok{\%>\%}\StringTok{ }
\StringTok{  }\KeywordTok{remove\_axes}\NormalTok{()}
\end{Highlighting}
\end{Shaded}

\begin{figure}[H]
\includegraphics[width=0.3\linewidth]{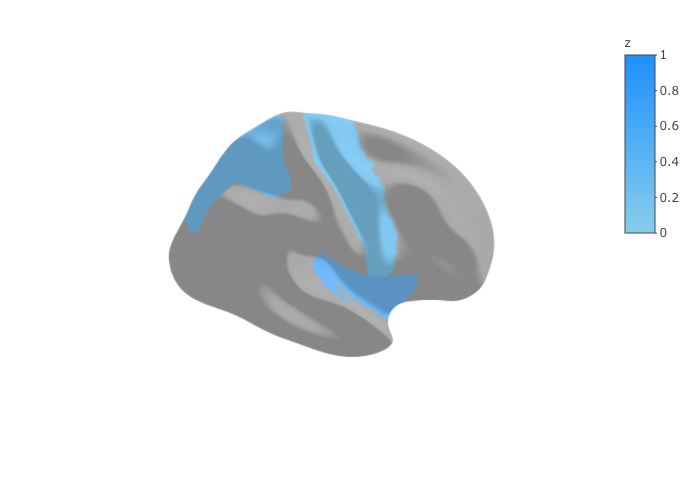} \includegraphics[width=0.3\linewidth]{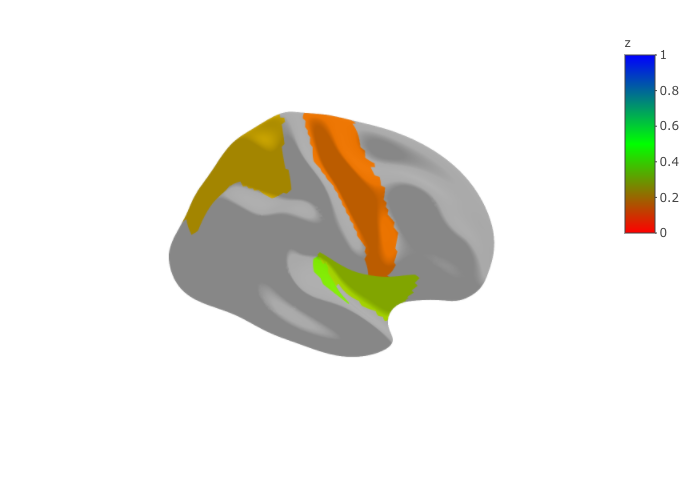} \includegraphics[width=0.3\linewidth]{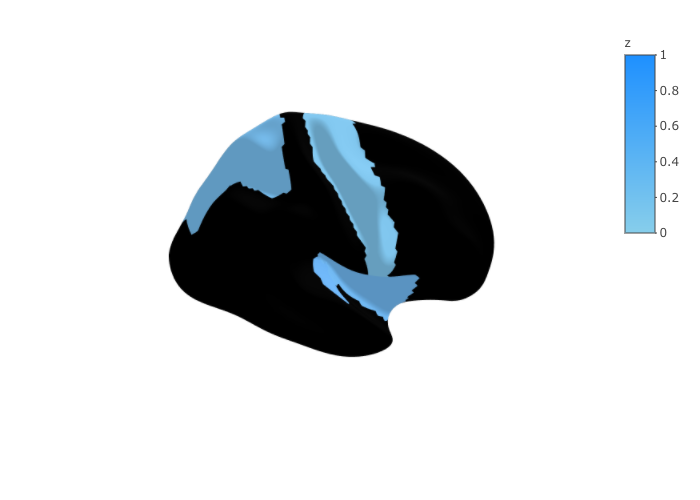} \caption{\textbf{Left:} Supplying data to \texttt{ggseg3d()} works similarly to \texttt{ggseg()}. Since \texttt{ggseg3d()} is based on \emph{plotly} rather than \emph{ggplot2}, most aesthetic adaptations must be set in the main function. Here we set the parcellation colours with the `colour' column and accompanying mouse-hover text with the `text' column. \textbf{Middle:} The palette for \emph{ggseg3d} needs to be specified directly in the main plot call. The `palette' option takes vectors of colours either as HEX-codes or R-colour names. \textbf{Right:} The colour of the \texttt{NA} values can also be changed through the option `na.colour'.}\label{fig:ggseg3d-data1}
\end{figure}

\hypertarget{customizing-colours-and-the-colour-bar}{%
\subsubsection{Customizing colours and the colour bar}\label{customizing-colours-and-the-colour-bar}}

You can provide custom colour palettes either in hex or R-names, as seen in Figure \ref{fig:ggseg3d-data1}.
Colours will be evenly spaced when creating the colour-scale.
A palette may also be supplied as a \emph{named numeric vector}, where the vector names are the colours that users wish to use, and the numeric values are the breakpoints for each colour (e.g.~\texttt{c("red"\ =\ 0,\ "white"\ =\ 0.5,\ "blue"\ =\ 1)}).
This way the users can control the minimum and maximum values of the colour scale, and also how the gradient is applied.
If another colour than the default gray is wanted for the \texttt{NA} regions, supply `na.colour', either as HEX colour or colour name.
This option only takes a single colour.

\hypertarget{adding-a-glass-brain}{%
\subsubsection{Adding a glass brain}\label{adding-a-glass-brain}}

\begin{figure}[H]
\includegraphics[width=0.6\linewidth]{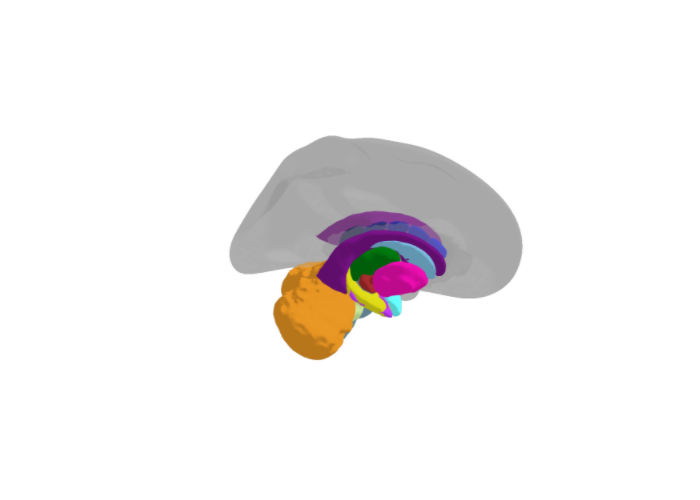} \caption{For subcortical structure visualization, one can add a glass brain to the plot. This will help with locating the structures relative to the cortex, and make the plot easier to interpret. The glass brain is controlled by three options: opacity, hemisphere, and colour.}\label{fig:glassbrain1}
\end{figure}

Subcortical atlases include cortical surfaces and other landmark structures for visualization purposes only.
One can control the opacity of the these \texttt{NA} structures, to improve visualization.
Glass brains can be added to provide a frame of reference for the subcortical structures with the function \texttt{add\_glassbrain()} (Figure \ref{fig:glassbrain1}) , which takes three extra arguments: hemisphere, colour, and opacity.

\begin{Shaded}
\begin{Highlighting}[]
\CommentTok{\# Figure 9}
\KeywordTok{ggseg3d}\NormalTok{(}\DataTypeTok{atlas =}\NormalTok{ aseg\_3d,}
        \DataTypeTok{na.alpha=} \FloatTok{.5}\NormalTok{) }\OperatorTok{\%>\%}\StringTok{   }
\StringTok{  }\KeywordTok{add\_glassbrain}\NormalTok{(}\StringTok{"left"}\NormalTok{) }\OperatorTok{\%>\%}\StringTok{ }
\StringTok{  }\KeywordTok{pan\_camera}\NormalTok{(}\StringTok{"left lateral"}\NormalTok{) }\OperatorTok{\%>\%}\StringTok{ }
\StringTok{  }\KeywordTok{remove\_axes}\NormalTok{()}
\end{Highlighting}
\end{Shaded}

\texttt{ggsed3d()} is based on \emph{plotly} and thus additional \emph{plotly} functionalities can be used to modify and improve the 3D atlas representations.
In addition to Carson Sievert's book on \emph{plotly} in R (\citeyearpar{plotly}), we recommend resources for modifying axes in 3D plots (\citet{plotly-ax}), the basic introduction to tri-surface plots (\citet{plotly-tri}), and this tutorial on tri-surface plots with \emph{plotly} in R (\citet{plotly-trisurf}).
Finally, we recommend \href{https://github.com/plotly/orca\#installation}{orca} command line tool to save \emph{ggseg3d} atlas snapshots.

\hypertarget{additional-atlases}{%
\subsection{Additional atlases}\label{additional-atlases}}

\begin{figure}[H]
\includegraphics[width=0.8\linewidth]{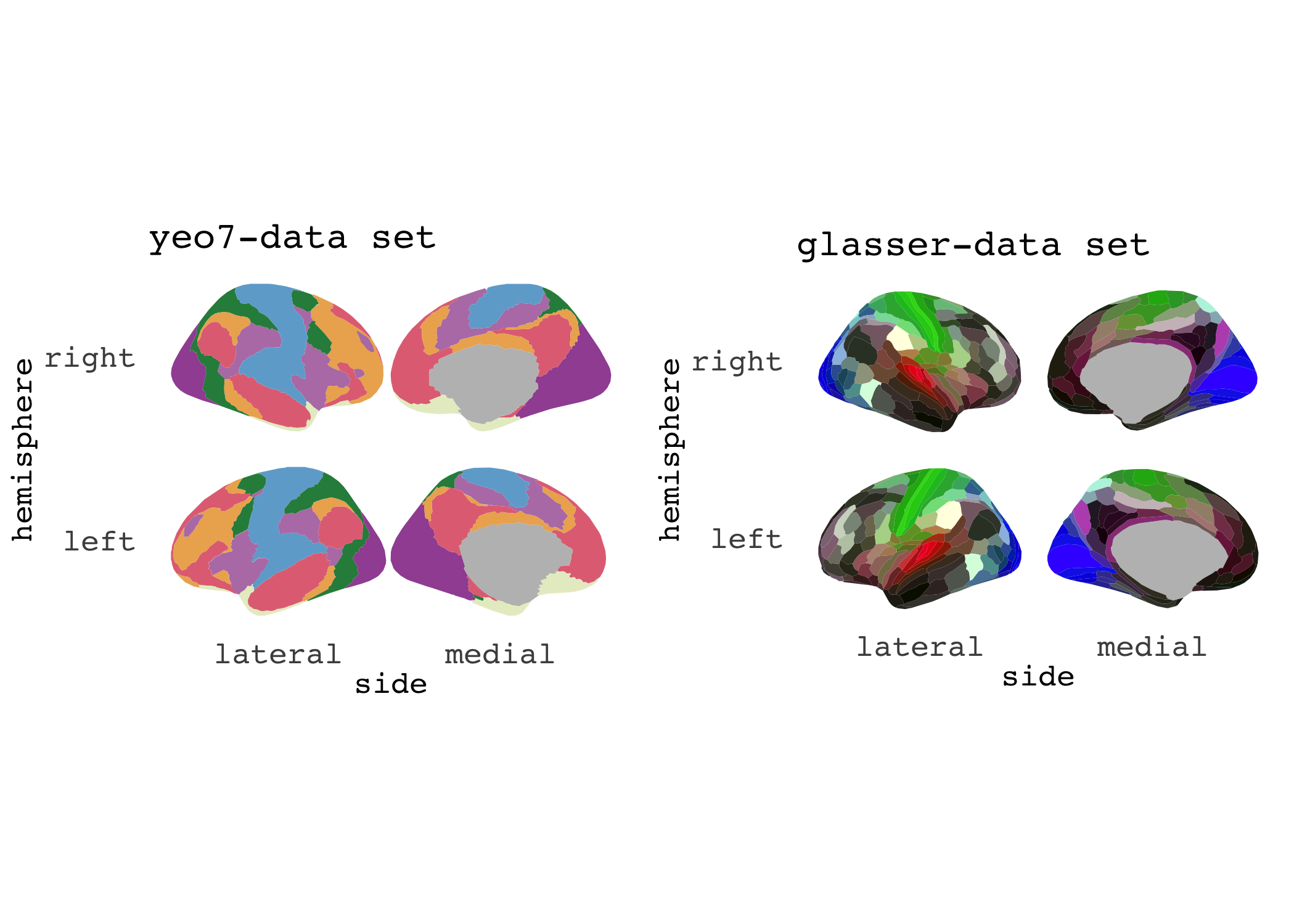} \includegraphics[width=0.4\linewidth]{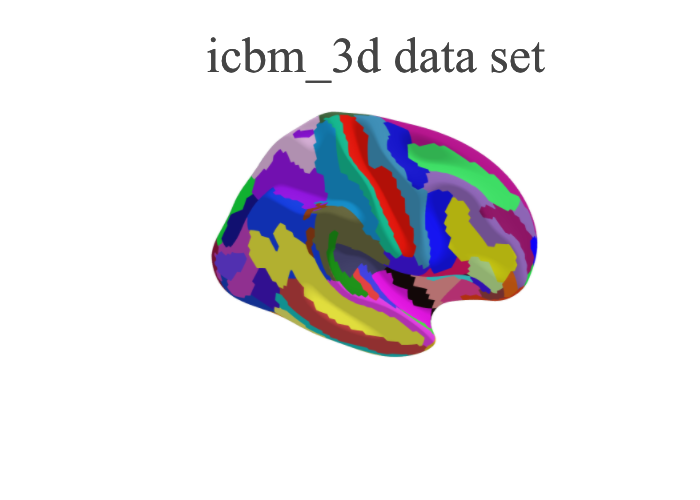} \includegraphics[width=0.4\linewidth]{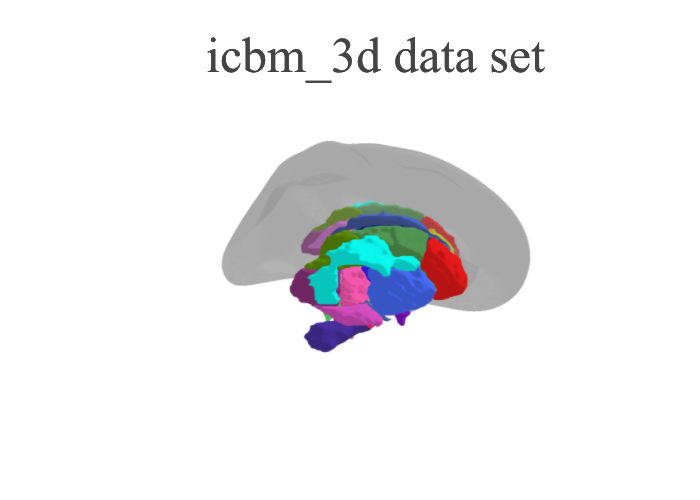} \caption{Four example data sets from the \emph{ggsegExtra} library, plotted with the \texttt{ggseg()} and \texttt{ggseg3d()} functions.}\label{fig:ggsegExtra}
\end{figure}

The \emph{ggseg} and \emph{ggseg3d} packages have two atlases each, which are 2D and 3D variations of the same main atlases: the \texttt{dkt} (\citet{dkt}) and \texttt{aseg} (\citet{aseg}) atlases.
These are, however, only two among many meaningful ways of segmenting the brain into different regions.
Thus, the \emph{ggsegExtra} package is a repository containing additional with additional data sets for plotting with the \emph{ggseg} and \emph{ggseg3d} packages.
There is an ever-increasing amount of new atlases being created, as research and methods in neuroimaging analysis progresses.
The \emph{ggsegExtra}-package is intended to be expanded as a community-effort, as new and informative atlases are published.
A small collection of the atlases currently in the \emph{ggsegExtra} package may be viewed in Figure \ref{fig:ggsegExtra}

\hypertarget{discussion}{%
\section{Discussion}\label{discussion}}

The main aim of the \emph{ggseg}, \emph{ggseg3d}, and \emph{ggsegExtra} packages is to ease and streamline visualization of brain atlas data in R, by gathering a collection of atlases from several scientific sources and providing customized plotting functions.
In this tutorial, we introduced the packages to the readers by presenting some use examples and highlighting the main functions and options that are available.
As visualization tools, these packages add up to manifold functionalities such as ggBrain (\citet{ggBrain}) and ggneuro (\citet{ggneuro}) in R, and software-specific image viewers such as FSLeyes (\citet{fsleyes}) and Freeview (\citet{dale_99}).
In this regard, we do not aim to compete with software-specific visualizations or advocate for the superiority of the \emph{ggseg}-packages as visualization tools.
After all, flattened 2D polygons do not rely on a meaningful brain coordinate system and the units of information in 3D meshes are limited to the number of parcellations.
On the contrary, we believe the \emph{ggseg} niche among visualization tools resides in its simplicity and its ability to be combined with statistical analysis pipelines.
The possibility to serve as an interactive tool for dissemination and reproducibility when combined with other technologies, such as Binder (\citet{binder}) or Shiny (\citet{shiny}), is an added benefit.
This is exemplified in the online supplementary information of Vidal-Piñeiro et al.~(\citeyearpar{vidal_2019}).

The three \emph{ggseg}-packages contain three main features:

\textbf{1)} a collection of 2D-polygon and 3D mesh brain parcellation atlases. The atlas data include the necessary coordinates for plotting, and include other information that should be recognizable for users.\\
\textbf{2)} \texttt{ggseg()} and \texttt{ggseg3d()} functions for visualization. Both functions are flexible and well-adapted to their environment and can be combined with any additional argument from \emph{ggplot2} and \emph{plotly}, respectively.\\
- \texttt{ggseg()} is a wrapper function for \texttt{geom\_polygon} from \emph{ggplot2} and it can be built upon like any \emph{ggplot2} object.\\
- \texttt{ggseg3d()} is a \emph{plotly} wrapper function for tri-surface mesh plots which prints 3D atlases.\\
\textbf{3)} Complimentary features -- e.g.~color scales - and functions such as \texttt{as\_ggseg\_atlas()} and \texttt{as\_ggseg3d\_atlas()} to convert data in the correct atlas format.

These functions provide users with the possibility of adapting the plots to their wishes, and also makes it possible to create and contribute to the atlas repository in \emph{ggsegExtra}.

The foundations of the \emph{ggseg}-packages trace back to the necessity of visualizing and exploring the lifespan trajectories of cortical thickness across different brain regions (see supplementary information in Vidal-Piñeiro et al.~(\citeyearpar{vidal_2019}) ) .
That is, \emph{ggseg} appears with the need to inspect and display brain information over time - i.e.~including a spatial dimension and a time-varying factor - overcoming the constrains of printed journals and classical 2D plots (e.g.~bar plots).
The current state of science requires researches to share the results of studies in both high detail and in an intuitive manner, as it permits communication to wide audiences and facilitates reproducibility.
Hence, we believe this tool conforms to the essence of open science and invite users to improve the code, provide examples, or tutorials, and contribute to the atlas collection according to their own interest and needs via the public \href{https://github.com/LCBC-UiO/ggseg}{ggseg GitHub repository}, \href{https://github.com/LCBC-UiO/ggseg3d}{ggseg3d GitHub repository}, and \href{https://github.com/LCBC-UiO/ggsegExtra}{ggsegExtra GitHub repository}.

Finally - while the \emph{ggseg}-packages are circumscribed to brain parcellations - we believe that the structure and functions of the package can be easily applied to any scientific field that benefits from data being displayed across the spatial dimension.
We encourage readers to borrow the package functionalities and adapt it to their respective fields and structures of interest, such as has already been done with the \emph{gganatogram}-package (\citet{gganatogram}).

\hypertarget{planned-package-improvements}{%
\section{Planned package improvements}\label{planned-package-improvements}}

In the \emph{ggsegExtra} \href{https://github.com/LCBC-UiO/ggsegExtra/wiki/Contributing\%3A-polygon-atlases-new}{github wiki}, we offer a pipeline to create and supply atlases for 2D plotting.
At the moment, the creation of atlas for ggseg is convoluted and difficult and requires manual intervention.
We are in the process of designing a simple, straightforward pipeline to facilitate the creation of new ggseg (2D) atlases.
The aim is to create a set of functions that will call specialized tools like FSL(\citet{fsl}), Freesurfer(\citet{fischl_99}, \citet{dale_99}, \citet{Fischl2000}) and Imagemagick(\citet{magick}) to detect the polygon vectors or the mesh segments given an MRI image containing a parcellation specification, and organize these into valid \emph{ggseg} and \emph{ggseg3d} atlases.
We encourage users to contribute to the \emph{ggsegExtra} brain atlas repository by including additional brain atlases.

\hypertarget{conclusion}{%
\section{Conclusion}\label{conclusion}}

Visualization is a fundamental aspect of neuroimaging to explore and understand data, guide interpretation, and communicate with colleagues and the general audience.
In this tutorial, we have introduced the \emph{ggseg}-packages, tools for visualizing brain statistics through brain parcellation atlases in R.
This visualization tool easily combines with interactive routines as well as with diverse statistical analysis pipelines.
We hope this tool and tutorial proves useful to neuroscientists and inspires others to apply the functions in a wide variety of fields and structures.

\hypertarget{author-contributions}{%
\section{Author Contributions}\label{author-contributions}}

Didac Vidal-Piñeiro generated the idea for the tool, and the initial scripts for plot visualization.
He has also been responsible for converting images from neuroimaging data to ggseg-like data (e.g.~polygons and mesh data).
Athanasia M. Mowinckel adapted the initial scripts and made the functions into package format and has continued developing the functions with the aim of increasing user-friendliness.
She is also responsible for conceiving and adding the mesh-plot functionality through plotly, and developing the pipeline for making that possible.
A. M. Mowinckel wrote the first draft of the paper, and both have since critically edited it.

\hypertarget{conflicts-of-interest}{%
\section{Conflicts of Interest}\label{conflicts-of-interest}}

The authors declare that there were no conflicts of interest with respect to the authorship or the publication of this article.

\hypertarget{acknowledgements}{%
\section{Acknowledgements}\label{acknowledgements}}

We thank John Muschelli for package code comments and its adaptation to neuroconductor ({[}2018{]}), and Richard Beare for the first community created atlas, and base code comments.
We are indebted to A. M. Winkler scripts for converting neuroimaging data (Freesurfer) into .ply files, a necessary step for neuroimaging to mesh-plots conversion, and to Inge Amlien for the `LCBC' surface files and helpful contributions to the ggseg3d pipeline.
We are indebted to those users that actively contributed to the package and all the individuals that worked on the framework on which the package is sustained (e.g.~R-project, neuroimaging software).
Finally, we thank Anders M. Fjell (AMF) and Kristine B. Walhovd (KBW) for their encouragement and financial support.

\hypertarget{funding}{%
\section{Funding}\label{funding}}

This work is funded by EU Horizon 2020 Grant `Healthy minds 0-100 years: Optimizing the use of European brain imaging cohorts (Lifebrain)', with grant agreement 732592.
The project has also received funding from the European Research Council's Starting (grant agreements 283634, to A.M.F. and 313440 to K.B.W.) and consolidator Grant Scheme (grant agreement 771355 to KBW and 725025 to AMF).
The project has received funding through multiple grants from the Norwegian Research Council"

\hypertarget{prior-versions}{%
\section{Prior versions}\label{prior-versions}}

Athanasia Monika Mowinckel also has several tutorials on her blog regarding \emph{ggseg} creation and functionality (\citet{ggsegAnim}, \citet{ggsegIntro}).

\renewcommand\refname{References}
\bibliography{references}
% \bibliography{msc_ggseg.bbl}

\end{document}